\documentclass[sigconf, nonacm=true]{acmart}
\setcopyright{none}
\AtBeginDocument{%
  }

\setcopyright{acmlicensed}
\copyrightyear{2025}
\acmYear{2025}
\acmDOI{XXXXXXX.XXXXXXX}
\acmISBN{978-1-4503-XXXX-X/2025/03}

\usepackage{cleveref}
\usepackage{amsmath}
\usepackage{algorithm}
\usepackage{algpseudocode}
\usepackage{tabularx}
\usepackage{todonotes}
 \usepackage{multirow}
 \usepackage{graphicx}
\usepackage{subcaption}
\usepackage{float} 
\usepackage{caption} 
\usepackage{gensymb}

\begin{document}


\title{On the Thermal Vulnerability of 3D-Stacked High-Bandwidth Memory
Architectures}


\author{Mehdi Elahi}
\orcid{0009-0008-1153-4071}
\email{melahi@aggies.ncat.edu}
\affiliation{%
  \institution{Department of Computer Systems Technology, \\North Carolina A\&T State University}
  \city{Greensboro}
  \state{North Carolina}
  \country{USA}
  }

  \author{Mohamed R. Elshamy}
\email{elshamy@nmsu.edu}
\orcid{0000-0003-1280-333X}
\affiliation{%
  \institution{Klipsch School of ECE, \\New Mexico State University}
  \city{Las Cruses}
  \state{New Mexico}
  \country{USA}
}

\author{Abdel-Hameed A. Badawy}
\orcid{0000-0001-8027-1449}
\email{badawy@nmsu.edu}
\affiliation{%
 \institution{Klipsch School of ECE, \\New Mexico State University}
   \city{Las Cruses}
  \state{New Mexico}
  \country{USA}
  }

\author{Ahmad Patooghy}
\orcid{0000-0003-2647-2797}
\email{apatooghy@ncat.edu}
\affiliation{%
  \institution{Department of Computer Systems Technology, \\North Carolina A\&T State University}
  \city{Greensboro}
  \state{North Carolina}
  \country{USA}
}

\renewcommand{\shortauthors}{Elahi et al.}

\begin{abstract}


3D-stacked High Bandwidth Memory (HBM) architectures provide high-performance memory interactions to address the well-known performance challenge, namely the memory wall. However, these architectures are susceptible to thermal vulnerabilities due to the inherent vertical adjacency that occurs during the manufacturing process of HBM architectures. We anticipate that adversaries may exploit the intense vertical and lateral adjacency to design and develop thermal performance degradation attacks on the memory banks that host data/instructions from victim applications. In such attacks, the adversary manages to inject short and intense heat pulses from vertically and/or laterally adjacent memory banks, creating a convergent thermal wave that maximizes impact and delays the victim application from accessing its data/instructions. As the attacking application does not access any out-of-range memory locations, it can bypass both design-time security tests and the operating system's memory management policies. In other words, since the attack mimics legitimate workloads, it will be challenging to detect.

\end{abstract}



\maketitle

\section{Introduction}
\label{sec:Intro}

The growing disparity between processor performance and memory access speed—commonly referred to as the memory wall—has become a critical bottleneck in modern computing systems \cite{wu2024removing}. As computational throughput scales rapidly across domains such as artificial intelligence (AI), high-performance computing (HPC), and data-intensive edge workloads, the demand for memory bandwidth has outpaced the capabilities of conventional memory technologies. High Bandwidth Memory (HBM), architectures enabled by 3D-stacking and interposer-based integration, has emerged as a key architectural solution to address this fundamental challenge.

\begin{figure*}[t]
    \centering
    \includegraphics[scale=0.68]{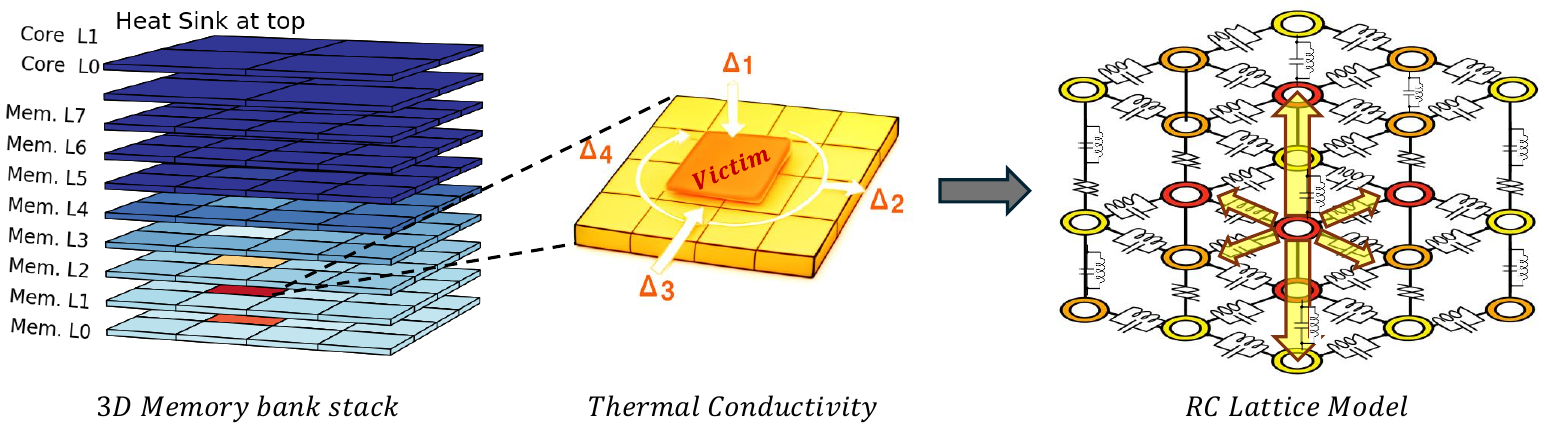}
    \caption{Thermal interaction map for a 4×4×8 HBM stack showing lateral coupling and vertical paths within and across layers} 
    \label{fig: Heat interaction}
\end{figure*}

HBM distinguishes itself from traditional memory designs through its deep stacking of dynamic random-access memory (DRAM) dies and its wide, parallel memory channel organization. As depicted in the left panel of \Cref{fig: Heat interaction}, each HBM stack integrates multiple dies interconnected using through-silicon vias (TSVs), forming vertically aligned banks that feed into a high-speed interface. These stacks are placed on a silicon interposer alongside compute units, allowing thousands of fine-pitch interconnects that deliver an order-of-magnitude improvement in memory bandwidth compared to off-package memory. The physical configuration—comprising wide I/O channels, multiple stack depths, and tightly coupled routing paths—enables high-bandwidth and low-latency data transfers critical for bandwidth-hungry workloads.

HBM offers massive throughput through increased stack depth and widened I/O channels, but its scalability introduces system-level challenges in power delivery, thermal management, and sustaining linear bandwidth across multiple stacks due to complex package-level integration \cite{lee2025thermal}. These issues must be addressed to fully harness next-generation HBM performance. Beyond memory research, HBM is critical in domains such as AI model training, where it accelerates tensor streaming in matrix multiplications; HPC workloads like climate modeling, physics, and genomics, which demand sustained bandwidth alongside compute throughput; and emerging edge computing platforms, where heterogeneous integration leverages HBM to balance real-time responsiveness with tight power constraints. As such, HBM is not only a key solution to the memory wall but also a foundational technology shaping future architectures across cloud, edge, and exascale systems through ongoing advances in organization, scalability, and system integration.

While HBM provides significant performance benefits, its 3D-stacked design also introduces degrees of architectural vulnerability stemming from vertical and lateral die adjacency, routing density, and thermal coupling. These aspects create non-trivial scaling challenges that remain insufficiently explored in current literature. In this work, we analyze these limitations in detail, offering new insights into how stack organization and interposer-level integration influence both performance scalability and reliability. Our contributions highlight critical design considerations that must be addressed to fully exploit HBM in next-generation AI, HPC, and edge computing platforms.

\section{HBM Vulnerability}

Heat propagation in 3D-stacked HBM is inherently anisotropic. Within a die, banks are located laterally on the same silicon layer, so heat primarily spreads in-plane, where the thermal conductivity of silicon is high and adjacent active regions are strongly coupled. Across dies, however, heat must traverse bonding interfaces, dielectrics, and TSVs, introducing substantially higher resistance than in-plane conduction. Consequently, lateral (intra-layer) heat spreading is faster and more effective than vertical (cross-layer) transport, and deeper stacks exhibit amplified temperature gradients. Layers farther from the primary heat sink---typically adjacent to the logic die---accumulate excess heat, while TSV placement and density further shape non-uniform vertical flow, influencing hotspot formation~\cite{lee2025thermal}.

This anisotropic behavior can be modeled by a compact RC thermal model depicted in~\Cref{fig: Heat interaction}- $"RC\_Lattice\_Model"$~\cite{meng2012optimizing} . Let $T_i(t)$ denote the temperature of bank $i$ and $P_i(t)$ its power input. Lateral heat transfer between adjacent banks $i$ and $j$ is modeled as shown in Eqn.~\ref{equ1}.
\begin{equation}
\label{equ1}
Q_{i \rightarrow j}^{lat}(t) = \frac{T_i(t) - T_j(t)}{R_{th}^{lat}(i,j)} \, ,
\end{equation}
where $R_{th}^{lat}$ reflects silicon’s in-plane conductivity and geometric adjacency. Similarly, vertical transfer between aligned banks $i$ and $k$ across layers is expressed in Eqn.~\ref{equ2}.
\begin{equation}
\label{equ2}
Q_{i \rightarrow k}^{vert}(t) = \frac{T_i(t) - T_k(t)}{R_{th}^{vert}(i,k)} \, ,
\end{equation}
with $R_{th}^{vert}$ generally much larger due to inter-die materials and interfaces. Aggregating all banks yields the standard RC network representation (Figure~\ref{fig: Heat interaction}), the head can be expressed as Eqn.~\ref{equ3}.
\begin{equation}
\label{equ3}
\mathbf{C}\,\frac{d\mathbf{T}(t)}{dt} + \mathbf{G}\,\mathbf{T}(t) = \mathbf{P}(t) \, ,
\end{equation}
where $\mathbf{C}$ is the diagonal thermal capacitance matrix and $\mathbf{G}$ encodes both the lateral and the vertical conductances. In practice, lateral conductances dominate within a die, while vertical pathways are bottlenecked by interface layers and TSV topology~\cite{agrawal2017xylem}.

Independent of specific memory-stack implementations, the fundamental imbalance between lateral and vertical conduction governs thermal behavior. Strong lateral coupling enables heat from active banks to quickly affect neighboring banks of the same layer, whereas relatively weak vertical conduction delays dissipation toward the heat sink. We project that the vertical and lateral proximity of HBM banks enables attackers to engineer thermal attacks that leverage convergent heat fluxes. In such attacks, a malicious actor injects short bursts of intense computational activity, effectively, heat pulses, into memory banks located adjacent (vertically and laterally) to those occupied by victim applications. Through careful coordination,
these pulses coalesce into a powerful thermal wave that, despite any additional activity, increases the temperature of the victim’s memory banks, thus activating preventive measures of thermal management, preventing access to critical data/instructions, and affecting the application
performance.

\section{Conclusions}
\label{Sec:ConC}

This work exposes a novel security vulnerability in 3D-stacked High Bandwidth Memory (HBM) systems, arising from the fundamental vertical and lateral adjacency inherent to their manufacturing process. By strategically orchestrating synchronized thermal pulses across vertically and laterally adjacent banks, our proposed methodology generates convergent thermal waves that degrade victim application performance while remaining virtually undetectable by conventional security and monitoring approaches. Through simulation-driven validation, we could demonstrate the feasibility and stealth of these thermal performance degradation attacks, underlining a critical new attack surface within contemporary memory architectures. Our findings highlight the urgent need for security mechanisms that address not just digital, but also physical and thermal interdependencies in emerging high-performance memory systems.



\balance
\bibliographystyle{ACM-Reference-Format}
\bibliography{References}
\end{document}